\documentclass[journal,twoside,web]{ieeecolor}
\usepackage{generic}

\usepackage[backend=biber,mincitenames=1,maxcitenames=1, maxbibnames=3, bibstyle=ieee, style=ieee, doi=false,isbn=false,url=false,eprint=false]{biblatex}
\PassOptionsToPackage{normalem}{ulem}
\usepackage{ulem}

\usepackage{amsmath,amssymb,amsfonts}
\usepackage{algorithmic}
\usepackage{graphicx}
\usepackage{textcomp}
\usepackage{gensymb}
\usepackage{color,soul}
\usepackage{lipsum}

\def\BibTeX{{\rm B\kern-.05em{\sc i\kern-.025em b}\kern-.08em
    T\kern-.1667em\lower.7ex\hbox{E}\kern-.125emX}}
\begin{document}

\title{Benchmarking real-time algorithms for in-phase auditory stimulation of low amplitude slow waves  with wearable EEG devices during sleep}

\author{Maria Laura Ferster, Giulia Da Poian, Kiran Menachery, Simon J. Schreiner,  Caroline Lustenberger, Angelina Maric, Reto Huber, Christian Baumann, and Walter Karlen, 
\thanks{ This project was conducted as part of the SleepLoop Flagship of Hochschulmedizin Zürich. We acknowledge funding from Tosoo AG and the ETH Foundation.}
\thanks{M.L. Ferster, G. Da Poian, K. Menachery, and C. Lustenberger are with the Department of Health Sciences and Technology, ETH Zurich, Switzerland. W. Karlen is with the Institute of Biomedical Engineering at the University of Ulm, Germany (email: walter.karlen@ieee.org). R. Huber is with the University Children’s Hospital Zurich and University Clinics for Child and Adolescent Psychiatry, University of Zurich. S. J. Schreiner, A. Maric, and C. Baumann are with the Department of Neurology, University Hospital Zurich, University of Zurich.  C. Baumann, R. Huber, W. Karlen, C. Lustenberger, A. Maric, and S. J. Schreiner,    are members of the Center of Competence \emph{Sleep \& Health Zurich}. 
}
}

\definecolor{added}{rgb}{0,0,0}
\newcommand{\added}[1]{{\color{added}{}#1}}

\maketitle

\begin{abstract}
Auditory stimulation of EEG slow waves (SW) during non-rapid eye movement (NREM) sleep has shown to improve cognitive function when it is delivered at the up-phase of SW. SW enhancement is particularly desirable in subjects with low-amplitude SW such as older adults or patients suffering from neurodegeneration such as  Parkinson disease (PD). However, existing algorithms to estimate the up-phase suffer from a poor phase accuracy at low EEG amplitudes and when SW frequencies are not constant.
We introduce two novel algorithms for real-time EEG phase estimation on autonomous wearable devices. The algorithms were based on a phase-locked loop (PLL) and, for the first time, a phase vocoder (PV). We compared these phase tracking algorithms with a simple amplitude threshold approach. The optimized algorithms were benchmarked for phase accuracy, the capacity to estimate phase at SW amplitudes between 20 and 60 \micro V, and SW frequencies above 1 Hz on 324 recordings from healthy older adults and PD patients. Furthermore, the algorithms were implemented on a wearable device and the computational efficiency and the performance was  evaluated on simulated sleep EEG, as well as prospectively during a recording with a PD patient.   
All three algorithms delivered more than 70\% of the stimulation triggers during the SW up-phase. The PV showed the highest capacity on targeting low-amplitude SW and SW with frequencies above 1 Hz. The testing on real-time hardware revealed that both  PV and PLL have marginal impact on \added{microcontroller load}, while the efficiency of the PV was 4\% lower than the PLL.
Active auditory stimulation did not influence the phase tracking. This work demonstrated that phase-accurate auditory stimulation can be delivered during home-based sleep interventions with a wearable device also in populations with low-amplitude SW.

\end{abstract}


\section{Introduction}
\label{sec:introduction}

\IEEEPARstart{T}{he} development of accurate algorithms for real-time electroencephalogram (EEG) phase detection is of high interest to a wide community, such as cognitive neuroscientists and rehabilitation engineers. The analysis of the EEG phase enables the decoding of different neuronal mechanisms and is used to understand the propagation and synchronization of neuronal firing \cite{Ng2013,Sauseng2008}. In addition, the EEG phase provides evidence regarding the exchange of information between global and local neuronal networks, and the sequential temporal activity of neuronal processes in response to external sensory stimuli. The real-time estimation of the EEG phase enables the state detection of those neuronal processes. For example, stimulating at specific phases during sleep amplifies slow waves (SW)  \cite{Ngo2013}, cognitive states are identified from the phase to exert control in brain-computer interfaces (BCI) \cite{VigueGuix2020}, or  neuronal processes  are modulated through external magnetic stimuli at specific phases \cite{Shirinpour2019}. Thus, real-time phase estimation algorithms tailored to the individual’s  brain oscillations are essential for interventions that require precise modulation.

The process of EEG phase estimation in real time faces several technical challenges. The estimation of the instantaneous EEG phase through non-causual and time-invariant methods such as the Hilbert transformation require data from before and after the event of interest and are therefore not suitable for real-time systems \cite{Chavez2006}. This challenge also applies to any filtering process that introduces non-linear phase delays \cite{Navid2019,Widmann2012}. In addition, real-time algorithms must be computationally efficient to minimize processing delays. Therefore, new strategies are needed to reduce application specific delays and lead to more accurate real-time phase estimation algorithms.  

A common application relying on real-time phase estimation is the auditory stimulation of brain activity during non-rapid eye movement (NREM) sleep. When stimulating during NREM sleep, the timing of the stimuli with respect to the EEG phase is crucial. Auditory stimulations applied in the up-phase of the SW (between 0$^{\circ}$ and 90$^{\circ}$) increase the SW amplitude
, whereas, stimulating in the down-phase (between 180$^{\circ}$ and 270$^{\circ}$) has no or a decreasing effect \cite{Ngo2013,Schabus2012}. Similarly, projecting these findings to cognitive outcomes, slow-wave activity (SWA) enhancement by up-phase SW stimulation resulted in an improvement of declarative memory  \cite{Leminen2017}, whereas targeting the down-phase did not \cite{Fattinger2017,Weigenand2016a}.

To limit the auditory stimulation to the up- or down-phase of SW, several real-time strategies have been introduced in the past years. The simplest is an amplitude threshold (AT) method without phase tracking \cite{Fattinger2019}. In this method, the EEG signal is band-pass filtered and auditory tones are triggered whenever the filtered EEG amplitude crosses a predefined threshold. Although this method has been shown to be effective \cite{Fattinger2019}, the non-linear phase delays\added{ introduced by the implemented real-time band-pass filter distort the signal and produce, unlike phase-true filters available in the post-processing analysis, a less than optimal  stimulation phase and amplitude accuracy.} In addition, the stimulations are limited to the amplitude characteristics of the SW. Consequently, low amplitude waves are not stimulated, and a predefined threshold can target a different phase depending on the characteristics of the SW \added{ (Fig. \ref{fig_challenge}(a))}. 
In 2016, the phase-locked loop (PLL) was introduced as an alternative method to estimate the SW phase during NREM sleep \cite{Santostasi2015}. On the given laboratory instrumentation, the PLL is sufficiently efficient and follows the temporal behavior of the EEG during slow-wave sleep (SWS). The internal frequency of the PLL is fixed around 1 Hz, such that it can follow EEG brain waves with a predominant frequency in the range of the SW frequency. However, a fixed frequency limits its capacity to follow oscillatory processes which are not close to 1 Hz. Consequently, this strategy does not lead to a purely in-phase stimulation, but a mixture that also includes rhythmical stimulation at 1 Hz when the PLL is not locked to the EEG signal frequency  \added{(Fig. \ref{fig_challenge}(b))}.

Existing methods for in-phase stimulation are challenged by altered sleep. In older adults and patients with neurological disorders, such as Parkinson disease (PD), sleep is hallmarked by a decrease of SWA \cite{Espiritu2008,Brunner2002,Latreille2015}. This decrement is an effect of reduced amplitudes and a lower number of SW compared to young healthy adults’ sleep characteristics. Previous studies report difficulties to enhance SWA in an older population during NREM sleep, showing only small or no effect  when comparing  stimulation and non-stimulation conditions \cite{Manuscript2018b,Papalambros2017,Schneider2020SusceptibilityAge}. When the PLL was used on older adults, a high spreading of phase at stimulation time was obtained \cite{Papalambros2017}. The presence of low amplitude SW indicates a decrease in the PLL accuracy with respect to targeting the SW up-phase. 
Consequently, the delivery of in-phase auditory stimulation for  populations with altered sleep requires specially designed algorithms that are also responsive to low SW amplitudes.

In an efficient system, real-time algorithms and adjacent software must be adapted to the specifications of hardware and the application needs. Sleep research studies, including the investigations on auditory stimulation that were introduced in the previous paragraphs, have traditionally relied on high-end EEG systems. These EEG systems are the gold standard for acquiring scalp EEG as they offer high signal quality and the possibility of mapping the brain activity to different brain regions. However, they can be expensive, and their use cannot be easily scaled to studies that aim for larger numbers of subjects over multiple nights. Small, head-mounted wireless EEG amplifiers have recently been introduced to investigate brain activity in real-life situations. Out-of-the-lab setups that acquire continuous EEG during everyday life are feasible, e.g. for ambulatory sleep monitoring or the diagnosis of neurological diseases \cite{Casson2018}. Amongst others, we recently showed that a portable biosignal processor can record high quality sleep EEG, represent the sleep macro- and microstructure in high detail, and perform autonomous auditory stimulation during NREM sleep \cite{Ferster2019}. In such wearable systems, the performance of the real-time EEG phase detection is crucial as no human supervision is possible. However, the specifications are even more stringent than in laboratory-based systems, as limited battery power and computational resources narrow down the range of technical solutions.

We have developed two novel algorithms to estimate the phase of SW with reduced amplitudes in real-time and address the limitations of fixed frequency stimulations and limited computational power of mobile devices. The goal of this work was to make algorithms more suitable for mobile applications, while increasing their phase tracking accuracy, especially in subjects with reduced SW activity during sleep. The first algorithm was a simplified version of the PLL that reduced the order of the previously published PLL architecture and consequently its computational complexity. The second algorithm introduced for the first time a strategy that relies on a phase vocoder (PV) architecture and was based solely on in-phase stimulation during NREM sleep adapting to the frequency changes of the EEG signal.   

In this work, we present the design and optimization process for the novel PLL and PV algorithms to provide real-time stimulation of SW up-phase during NREM sleep embedded on a wearable platform. We present a systematic evaluation of their phase precision and stimulation capacity on low and high SW amplitude and compare these to a simple AT approach without phase tracking. In addition, we compare the capacity of all algorithms on following SW frequency changes in real time. This benchmark is performed on a unique dataset that aims to represent populations with decreased SW amplitudes and fragmented NREM sleep. In addition to this benchmarking, we compare the efficiency between the PLL and PV when embedded on hardware. 
To validate the implementation on a resource-constrained, battery-powered wearable device and demonstrate the translation to users with altered SW, we prospectively evaluate the performance of the system on a patient with PD.These experiments illustrate that wearable, low-power devices can be used to deliver phase-accurate auditory stimuli during sleep also in older adults, as well as neurodegenerative patients \added{who} express SW with lower amplitudes.

\section{Algorithms}
\label{sec:algorithms}

 The in-phase auditory stimulation was performed with algorithms at several stages, interacting in real-time, and designed to run on the embedded microcontroller unit (MCU) of a wearable EEG device. 
The autonomous auditory stimulation algorithm was responsible for conditioning the delivery of the stimulation triggers throughout the night based on the recorded EEG signal. The aim of this algorithm was to prevent undesired stimulation in shallow sleep and monitor arousals and awakenings, achieved in different stages. The stage of main interest for this work was the SW phase estimation which was responsible to ensure that the auditory triggers could reliably target the desired SW up phase.
We aimed at re-innovating the strategies for real-time EEG phase estimation.  The baseline consisted of a trivial AT method that did not track phase. We additionally proposed two novel approaches: a first-order PLL that replicated a widely used phase tracking strategy, and a PV based approach that was novel to the field. All strategies relied on a common input with a single EEG derivation from the prefrontal (Fpz) - mastoid (M2) that was preprocessed with a) a 50 Hz notch filter to eliminate power line noise, b) a first order high-pass filter at 0.1 Hz to attenuate the DC component, and c) a first order low-pass filter at 30Hz to attenuate high frequency components.

\subsection{Autonomous auditory stimulation}
We implemented an autonomous auditory stimulation algorithm that consisted of a combination of sub-stages that processed a single EEG channel to generate an auditory stimulation trigger \cite{Ferster2019}.
A decision logic triggered the auditory stimuli when NREM sleep, SWA, EEG beta power, and EEG phase target conditions were simultaneously met. The NREM sleep detection classified 4 s of EEG signal into NREM sleep (sleep stages N2 and N3) or not-NREM sleep (awake, N1, and REM) based on the spectral power at different frequency bands (0.5 - 2 Hz, 2 - 4 Hz, and 20 - 30 Hz) from the past 80 s of EEG. If the power crossed predefined thresholds, the EEG signal was classified as NREM sleep. The SWA detection algorithm calculated the delta power (0.5 - 4 Hz) over the last 4 s. When delta power reached a predefined SWA threshold, the SWA condition was satisfied. A rise in beta power is indicative of awake, light sleep, artifacts, and the presence of arousals \cite{Berry2017}. Therefore, the EEG beta power detection inhibited triggers when high beta power (17-22 Hz) was detected in the last 4 s of EEG. 
For this work, all the thresholds of the previously mentioned stages were configured to have high trigger accuracy during NREM sleep for which we used sleep data from 11 healthy subjects (41 $\pm$ 16.1 years old, 6 f). The remaining EEG phase targeting stage  EEG phase was composed of  algorithms that were subject of investigation and are described in  detail in the following section. 

\subsection{Slow wave phase estimation}
\subsubsection{Amplitude threshold}
Fattinger et al., proposed a threshold-based approach where pink 1/f noise of $~$50 dB was played for precisely 50 ms as soon as a band-passed filtered EEG signal (Butterworth, 0.5 – 2 Hz) crossed a 50 \micro V EEG AT \cite{Fattinger2019}. Simultaneously, the experimenter continuously monitored the electromyography (EMG) information and adapted a second EMG threshold to decide whether to play the tones or not. We adopted this approach to an automatic real-time estimation of the EEG amplitude by relaxing the required pre-processing filtering condition that simultaneously reduced non-linear phase delays. The EMG condition was replaced by the auditory stimulation sub-stages presented in the previous section that were solely based on the EEG signal. The AT was the only parameter to optimize. 

\subsubsection{First-order phase-locked loop}
\begin{figure}[t]
\centerline{\includegraphics[width=0.8\columnwidth]{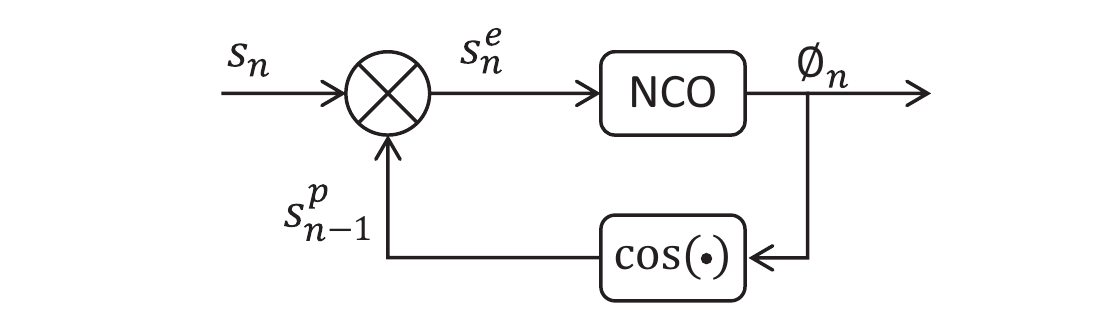}}
\caption{Phase-locked loop (PLL) block diagram. The EEG input signal $s_{n}$ is multiplied by the PLL output signal $s_{n-1}^p$ generating an error signal $ s_{n}^e$ that serves as a control signal for the number-controlled oscillator (NCO). To lock in phase, the NCO output signal minimizes the phase error with respect to $s_{n}$. Finally, $s_{n}^p$ is constructed as a cosine with a fixed frequency set at 1 Hz and phase $\phi_{n}^p$.}
\label{fig_PLL}
\end{figure}

The simplified PLL designed for this work consisted of a phase detector element and a number-controlled oscillator (NCO) element (Fig. \ref{fig_PLL}). The phase detector multiplied the prepossessed EEG signal $s_{n}$ with the PLL output $s_{n-1}^p$, which were characterized by sinusoidal and cosine functions, such as 

\begin{equation}
    s_{n} = A_{n}\sin(\omega_{n}+\phi_{n})
    \label{eqPLL1}
\end{equation}
\begin{equation}
    s_{n-1}^p = \cos(\omega_{n-1}^p+\phi_{n-1}^p)
    \label{eqPLL2}
\end{equation}

The phase detector output resulted in a signal error term \(s_{n}^e\) containing a high and low frequency component, such as

\begin{equation*}
    s_{n}^e = s_{n}s_{n-1}^p = \frac{A_{n}K_{\text{pd}}}{2}\sin(\omega_{n}+\omega_{n-1}^p+\phi_{n}+\phi_{n-1}^p) +
\end{equation*}
\begin{equation}
    \frac{A_{n}K_{\text{pd}}}{2}\sin(\omega_{n}-\omega_{n-1}^p+\phi_{n}-\phi_{n-1}^p)
    \label{eq:error}
\end{equation}
where $K_{\text{pd}}$ was the phase detector gain. 

To phase-lock with endogenous slow-wave oscillations, the PLL NCO frequency was set to 1 Hz. The simplified PLL did not use an additional low-pass filter to eliminate the high-frequency components. Different from typical PLL applications in the telecommunication field where dominant frequencies are in the order of GHz, the SW dominant frequency oscillated around 1 Hz and therefore the low and high frequency components are very close. Consequently, the low- and high-frequency components could not be separated without distorting the desired low-frequency component.

When the EEG signal frequency oscillated around 1 Hz, the $ \omega_{n}-\omega_{n-1}$ $\approx0$. We assumed small phase errors leading to
\begin{equation}
    \sin(\omega_{n}-\omega_{n-1}^p+\phi_{n}-\phi_{n-1}^p) \approx \phi_{n}-\phi_{n-1}^p
    \label{eq:phaseerror}
\end{equation}

The error signal $s_{n}^e$ was used as the NCO control signal. The NCO generated a periodic signal with a frequency that was proportional to $s_{n}^e$ and minimized the phase error with respect to $s_{n}$. The rate of change of the frequency in the NCO was given by its gain $K_{ \text{NCO}}$ and represented the sensitivity. Considering that the changes in frequency and phase from sample to sample did not variate drastically in the presence of SW, the sample-to-sample difference of the high-frequency term was neglected. Finally, the NCO output was constructed as a cosine signal with an angular frequency at 1 Hz and a phase given by $\phi_{n}^p = \phi_{n-1}^p - s_{n}^e K_{\text{NCO}}$.

The stimulation trigger was delivered when $\phi_{n}^p$ crossed a predefined target phase $\phi_{T}$.

The PLL gains $K_{\text{pd}}$ and $K_{\text{NCO}}$ were multiplied and  considered as one unique gain $K_{\text{PLL}}$. During optimization we tuned the $K_{\text{PLL}}$ and $\phi_{T}$ configuration parameters.

\begin{figure}[t]
\centerline{\includegraphics[width=0.9\columnwidth]{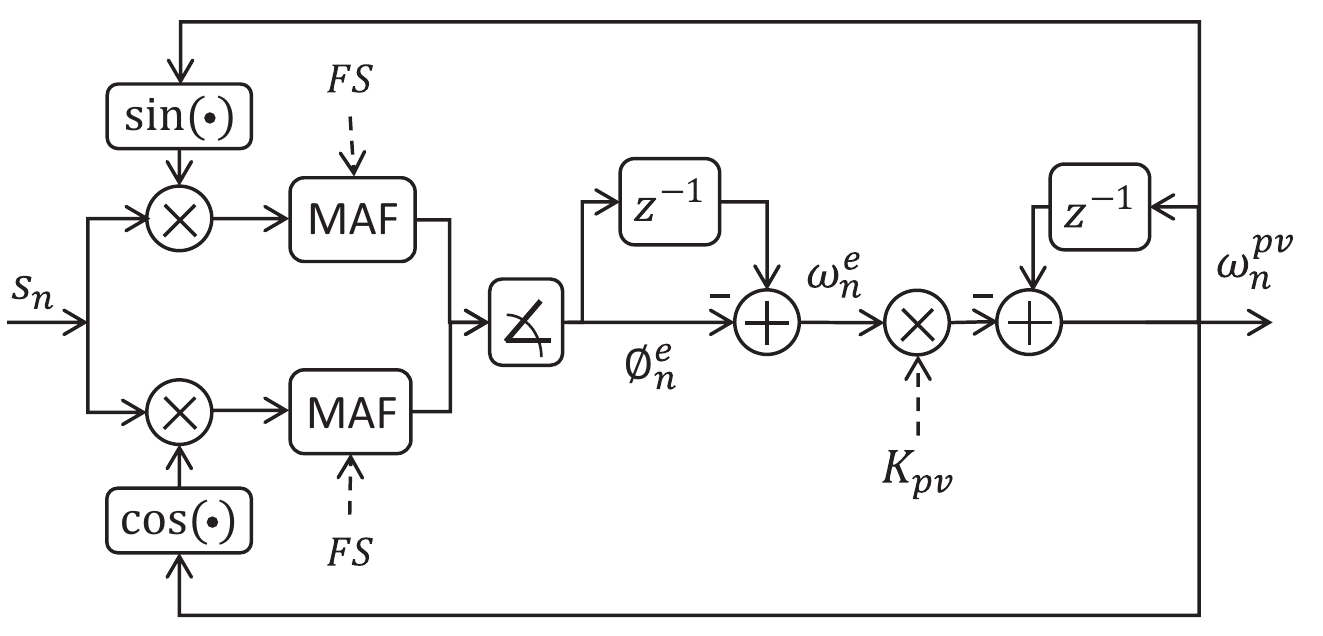}}
\caption{Phase vocoder (PV) block diagram. The EEG input signal $s_{n}$ is multiplied by a sine and cosine wave containing the PV argument $\omega_{n}^{pv}$. A moving average filter (MAF) eliminates the high frequency components. The two MAF outputs are converted from Cartesian to Polar coordinates to extract the phase error $\phi_{n}^e$, which is used to calculate the angular frequency error $\omega_{n}^e$ to finally update $\omega_{n}^{pv}$.}
\label{fig_PV}
\end{figure}

\subsubsection{Phase vocoder}

A PV uses frequency-domain transformations that are well-known to generate a variety of digital audio effects \cite{Dolson1986}. Originally introduced in the telecommunication domain as a military encoding machine, it quickly expanded to multiple audio and music applications. More recently, the PV found its way to biomedical time-series processing, i.e. for tracking heart rate in the photoplethysmogram \cite{Karlen2011} or detecting seizures in neonatal EEG \cite{Temko2014}. Therefore, we explored the PV for tracking the phase of EEG at its dominant frequency component. 

In the architecture of the PV the EEG input signal was routed into two parallel channels (Fig. \ref{fig_PV}). In the first channel, the signal was multiplied with a sine wave, and in the second channel with a cosine wave. The sine and cosine waves had unitary amplitude and angular frequency $\omega_{n-1}^{PV}$, corresponding to the estimated PV argument at the previous iteration. Both channels carried the same EEG information with a 90$^\circ$ shift in phase. Next, both signals were low-pass filtered with a moving average filter (MAF). Input frequencies in the proximity of the PV frequency were down-shifted close to DC to pass the MAF. All other frequency components were also shifted but did not pass the MAF. The length of the MAF  $FS$ determined how many past input samples were considered for the averaging. Next, the two filtered signals were transformed from Cartesian to Polar coordinates to calculate the phase error $\phi_{n}^e$ between the input signal and the PV. The angular frequency error $\omega_{n}^e$ was estimated as the difference between $\phi_{n-1}^e$ and $\phi_{n}^e$ and it was multiplied by a gain $K_{\text{PV}}$ to emphasize the sensitivity of the PV to the EEG frequency changes. Finally, the PV used the $\omega_{n}^e$ to update its angular frequency with respect to the input EEG signal.  

When the PV argument crossed a predefined target phase $\phi_{T}$, the stimulation was enabled. We optimized the PV for the $FS$, $K_{\text{PV}}$ and $\phi_{T}$ configuration parameters.

\section{Materials and methods}
\label{sec:MaterialAndMethods}

To systematically and objectively evaluate the three proposed SW phase estimation algorithms, we optimized each of their configuration parameters on a software simulator using an identical cross-validation setup and compared their performance on a test data set. In a second experiment, we assessed their efficiency on hardware. The two phase tracking algorithms (PLL and PV) were translated into an embedded system and the EEG signals from the test data set were fed to the device using a physical EEG simulator. Finally, in a third experiment we prospectively validated the algorithms' \added{performance} under realistic conditions in a human subject suffering from PD.

\subsection{Algorithm benchmarking}

\subsubsection{Data sets}
The purpose of the benchmarking was to challenge the algorithms with sleep EEG data that is not evident for phase tracking. Sleep micro and macro structures can present high inter- and intra-subject variability, which can be more pronounced in subjects with neurological diseases \cite{Petit2004}. To account for this, it was essential to optimize the algorithms on a very broad data set that reflected this variation. Therefore, we selected sleep data from PD patients to optimize the parameters, hypothesizing that the PD data set would also generalize to a healthy population with less variability in the sleep structure, which we verified on a second dataset. 

We extracted EEG data from 45 recordings from 14 subjects (56.1 $\pm$ 11.7 years old, 8 males) 
who participated in a calibration and usability study for the Mobile Health Systems Lab Sleep Band version 2 (MHSL-SBv2), a portable and configurable EEG system developed for mobile sleep research \cite{Ferster2019}. All subjects were diagnosed with PD.  They self-applied the system independently at home after a training session. Auditory stimulation was activated in 26 of the 45 recordings. For this work, we used the EEG derivation from the Fpz-M2 channel at a sampling rate of 250 Hz. Additionally, reference sleep scoring according to the AASM criteria \cite{Berry2017} with \added{a modified epoch length of 20 s} was available.  This scoring was performed by  sleep experts based on the same Fpz-M2 channel and additionally electrooculogram  (EOG)  and EMG  channels. 

 We  randomly selected 5 out of the 45 nights from 5 different subjects with PD  (54.4 $\pm$ 7.5 years old, 2 males) for the {\it PD test set} which was used to  benchmark the performance.  For the {\it optimization data set}, we used the remaining 40 nights. 

To test the generalization of our approach with independent data, we built a second, larger  {\it elderly test set} composed of 319 recordings of sleep EEG without stimulation from 24 healthy adults older than 62 y (68 $\pm$ 4.7 years old, 10 females), who used the MHSL-SBv2 during 14 consecutive nights as part of a larger clinical trial (NCT03420677) \cite{Lustenberger2021}. For the analysis of the {\it elderly test set}, sleep was automatically scored post-recording into slow-wave NREM (N2 and N3 sleep stages) and non-NREM sleep (N1, REM and awake) for 20-s long epochs by a validated deep-learning algorithm (unpublished). 

The overall analysis of this work considered  the N2 and N3 sleep stages characterized by SW  and sleep spindles as NREM sleep and  awake, N1, and REM sleep as not-NREM sleep. 

All the data used for for the optimization and benchmarking work was obtained from studies that were performed according to the Declaration of Helsinki and  approved by the responsible ethical review boards. 

\subsubsection{optimization}
We optimized the algorithms’ parameters based on three objectives: (1) the minimization of the mean error with respect to 45°, (2) the maximization of the stimulation percentage in the SW up-phase, and (3) the minimization of the stimulation percentage outside the SW up-phase range. Therefore, to find the parameter set that satisfied all three objectives, we used a multi-objective optimization method based on the minimum Euclidean distance (ED) with respect to the utopian solution. 

As outlined in the introduction, previous work highlighted the importance of targeting the SW up-phase to induce a SWA enhancement effect \cite{Zhang2019}, whereas targeting a down-phase would lead to a rather negative effect. Therefore, the first optimization objective focused on minimizing the circular mean absolute error (CMAE) with target angle 45° such as
\begin{equation}
    \text{CMAE}_{45} = \frac{|\text{circularMean}(\phi)-45^{\circ}|}{180^{\circ}}
\end{equation}

where  $\phi$ represented all the phases at stimulation trigger. The 45$^{\circ}$ target phase was selected to account for preprossessing filtering and processing delays.

Furthermore, it was of high interest to maximize the number of stimuli delivered within a night. The number of possible stimulations depended on individual sleep characteristics and varied strongly across nights. Therefore, optimizing for absolute numbers of stimuli would have lead to strong bias towards longer nights and those with longer continuous NREM sleep stages. For the second  optimization objective to be night and subject independent, we introduced a new metric that measures the percentage of active stimulations (PAS) in the SW up-phase $\text{PAS}_{\in\text{UP}}$ during NREM sleep, such as
\begin{equation}
\text{PAS}_{\in\text{UP}} = \frac{\#\, \text{of  Stimulations} \in (0-90^{\circ})}{(\#\, \text{of NREM windows})*(\text{MaxStim})}
\end{equation}
where the NREM windows were 2 s long of scored as NREM sleep that coincides with the device internal NREM sleep and SWA detection. To avoid multiple stimulations in the same SW we limited the stimulation frequency to 4 Hz. Thus, the maximum possible number of stimulation  in a 2 s window $MaxStim$ was 8.   

Finally, to avoid the selection of configurations that satisfies the two previous described objectives at a cost of high number of stimulation in the undesired SW phases, we included a third objective to minimize the PAS on those phases:
\begin{equation}
\text{PAS}_{\notin\text{UP}} = \text{PAS}_{ALL} - \text{PAS}_{\in\text{UP}}
\end{equation}
All optimization objectives were normalized from 0 to 1. As we aimed to minimize the CMAE$_{45}$ and $\text{PAS}_{\notin\text{UP}}$ and maximize the $\text{PAS}_{\in\text{UP}}$, the utopian solution ($\text{CMAE}_{45}$, $\text{PAS}_{\notin\text{UP}}$, $\text{PAS}_{\in\text{UP}}$) took the (0,0,1) coordinate in the normalized solution space. Therefore, we selected an algorithm’ parameters combination with the minimum Euclidean distance ($ED$) to (0,0,1). Consequently, we defined the $ED$ such as
\begin{equation}
\text{ED}=\sqrt{\text{CMAE}_{45}^2+\text{PAS}_{\notin\text{UP}}^2+(1-\text{PAS}_{\in\text{UP}})^2}
\end{equation}

\begin{figure}[!t]
\centerline{\includegraphics[width=\columnwidth]{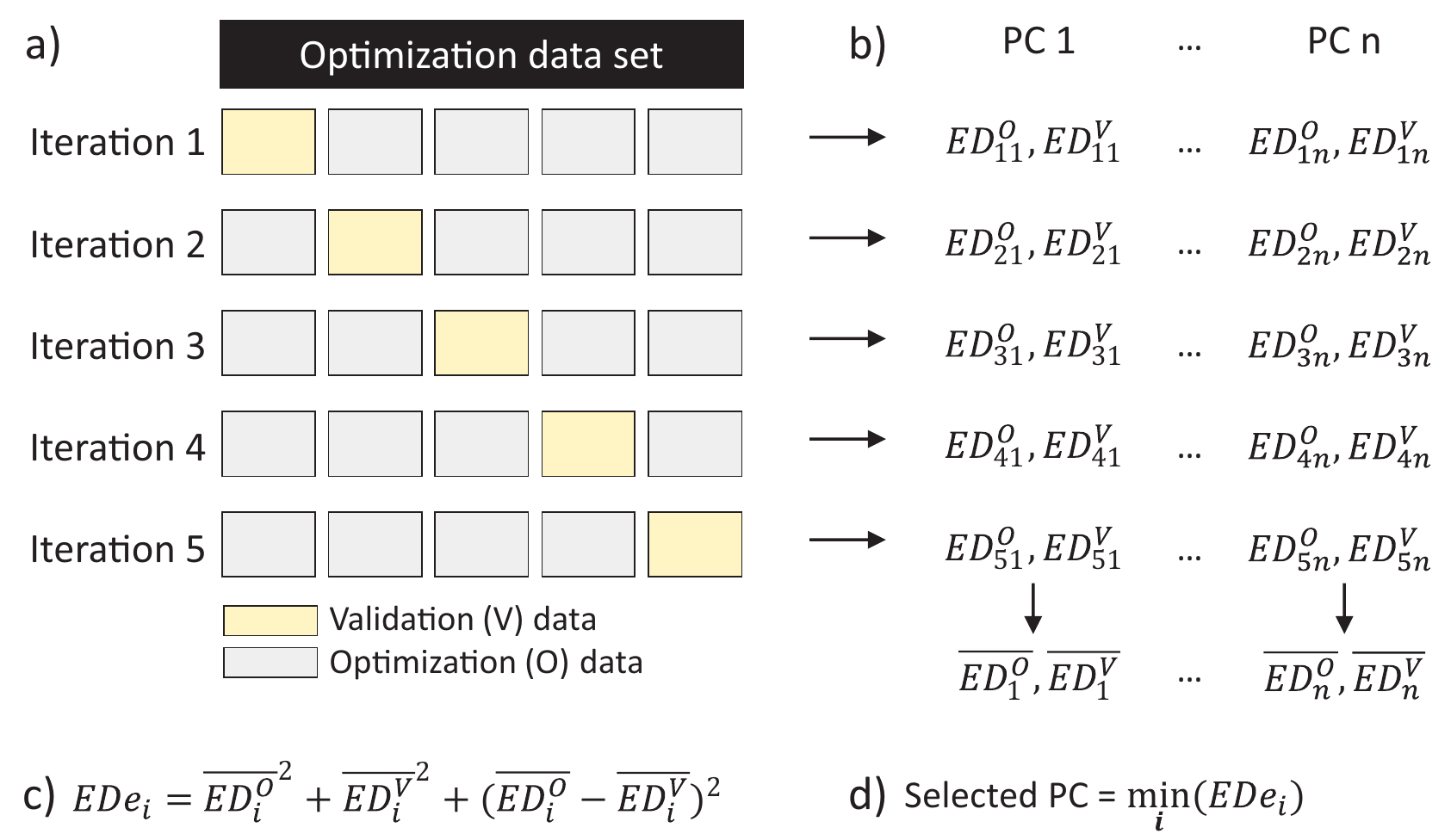}}
\caption{Cross-validation process for the algorithms’ parameters configuration (PC) selection. (a) PD optimization set divided in 5 equal sized folds (4 folds optimization, 1 fold validation). (b) Calculation of Euclidean distance (ED) for the n algorithm’s parameter combination (PC) on optimization and validation and ED average across iterations. (c) Calculation of ED error. (d) Selection of PC through the minimum ED error criteria.}
\label{fig_CV_steps}
\end{figure}

To generalize the algorithms’ parameters and avoid over fitting, we performed a K-fold cross validation on the optimization data set independently for each algorithm. We randomly divided the optimization data in 5 equally sized folds. Four out of 5 folds were used for optimization and 1 fold was used for validation (Fig. \ref{fig_CV_steps}(a)). For each iteration j, we calculated the ED for each algorithm’s parameter combination ($\text{PC}_{i}$) over the optimization ($\text{ED}_{ji}^{O}$) and validation ($\text{ED}_{ji}^{V}$)  data (Fig. \ref{fig_CV_steps}(b)). At the completion of all iterations, we calculated the ED average across iterations for each PC ($\overline{\text{ED}_{ji}^{O}}$,$\overline{\text{ED}_{ji}^{V}}$) followed by the calculation of the ED error for each PC (Fig. \ref{fig_CV_steps}(c)).
For benchmarking and further developments, we selected the PC with the minimum ED error over all folds (Fig. \ref{fig_CV_steps}(d)), therefore ensuring maximal generalizability and high performance.

\subsubsection{Data analysis and statistics}
 The ground-truth phase was \added{post-hoc}  extracted using the Hilbert transform applied over \added{the zero-phase band-pass filtered EEG signal (0.5 – 4 Hz,  Chebyshev type II) cropped by 210 s on both sides. We extracted the Hilbert phase at the timestamps where stimulation triggers were generated. }

We quantified the algorithms' ability to target the SW up-phase by calculating the accuracy with respect to the desired 45$\degree$   by calculating the  circular mean (CM) and standard deviation (CSD), CMAE$_{45}$, PAS for all triggers $\text{PAS}_{\text{ALL}}$, and  PAS for triggers presented in and outside the SW up-phase region ($\text{PAS}_{\in\text{UP}}$ and $\text{PAS}_{\notin\text{UP}}$). We depicted the trigger phase distribution of the entire data set for each algorithm in a circular histogram with 36 bins of 10 degrees and a mean vector. Statistical analysis was performed over the {\it PD test set} and across nights with paired sample t-tests using Bonferroni correction.

We evaluated the generalization capability of the algorithms to track the phase in an unseen population that had a different sleep structure.  We calculated the CMAE$_{45}$, CM, CSD, and all PAS for all nights and compare the resulting performance with the {\it PD test set} based on the objectives described in the optimization section. 

We compared the capacity of all algorithms in (1) targeting low and high amplitude SW, and (2) following the SW frequency changes above 1 Hz in real time. For this purpose, we extracted the amplitude and frequency characteristics of all waves from the filtered signal in the 0.5 to 4 Hz frequency band that coincided with the scored and the device NREM sleep classification. We defined the wave frequency as the inverse of the time between two minima and the wave amplitude as the peak-to-peak amplitude between the first minimum and the maximum of the detected wave. We determined the capacity of each algorithm to target the low (from 20 to 60 \micro V) and high (above 60 \micro V) SW amplitudes as the percentage of stimulated SW in each group. Finally, we compared the real-time algorithms' capacity on tracking the whole SW frequency spectrum band by extracting the triggers interval.

 All analysis was performed using Matlab 2018a (Mathworks Inc, USA). The CircStat toolbox \cite{Berens2009a} was used to calculate the circular statistics for the EEG phase.

\subsection{Translation to wearable hardware}
\begin{figure}[!t]
\centerline{\includegraphics[width=\columnwidth]{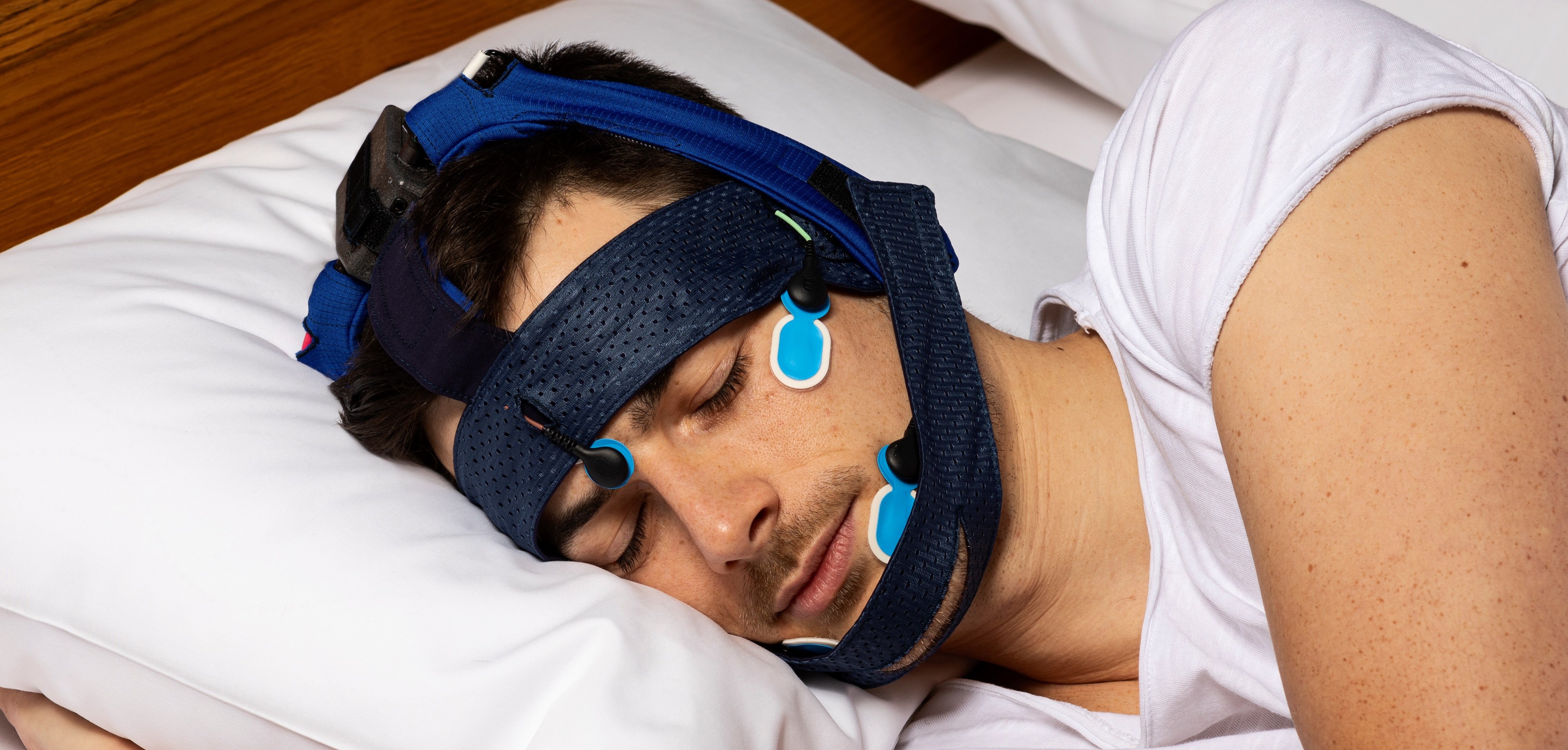}}
\caption{Wearable EEG, EOG and EMG recording device for autonomous auditory NREM sleep stimulation   (Tosoo AG, Switzerland).} 
\label{fig_Sleeploop}
\end{figure}

To assess the efficiency of the PLL and PV algorithms on wearable hardware, the algorithms were translated to a real-time embedded operating system (ConcertOS, Leitwert GmbH, Zurich, Switzerland) designed for running power constraint real-time applications on a wearable device (Tosoo AG, Zurich, Switzerland, Fig. \ref{fig_Sleeploop}). The Tosoo Axo A/B is more integrated version of the MHSL-SBv2 with better autonomy and internet-of-things connectivity. It features a 8-channel biosignal amplifier and analog-to-digital converter (ADS1299, Texas Instruments, USA) and a system-on-chip ARM-Cortex M4 ultra low-power MCU  (nRF52840, Nordic Semiconductor, Oslo, Norway). A 16 GB microSD card and a 660 mAh Li-Po battery (Varta,  Ellwangen, Germany) provided a recording autonomy of up to 24 h. Data was transmitted via WiFi securely to a remote server. The EEG time-series and the parameters calculated by the algorithm sub-modules, including all stimulation triggers, were recorded at biosignal sampling resolution.

To create a realistic workload for the system, the EEG recordings from the {\it PD test set} were fed to the Tosoo Axo A/B using an EEG waveform simulator. 
EEG signals were first processed using a notch  (zero-phase first-order IIR, 50 Hz) and high-pass filter (zero-phase first-order IIR, 0.1 Hz) to remove line noise and DC offset, and amplified (gain=5000) to improve the dynamic range.
A smartphone replayed the EEG signal simultaneously to two Tosoo Axo A/B devices running one of the phase detection algorithms each.  For this, the signal passed through the USB audio output and a custom built voltage divider \cite{Petersen2016}.

In addition to the performance metrics already introduced in the algorithm benchmarking section, we assessed the algorithms' efficiency on the target device. The  required  MCU cycles dedicated to algorithm processing
has a direct effect on battery autonomy and consequently indicate the algorithms' efficiency. We defined the required cycles rate (RCR) as the ratio between the median number of MCU cycles required by each algorithm and the total number of MCU cycles available between two EEG samples. We then calculated the efficiency such as \begin{equation}
\text{efficiency} = 100 * (1 - RCR).
\end{equation}
\added{In addition, we reported the 1$^{st}$  and 3$^{rd}$ quartiles (Q1, Q3) of the MCU cycles.}

\subsection{Translation to patients}
To investigate the algorithms' response to phase and frequency changes due to active auditory stimulation, we prospectively tested them in one PD patient (59 y, female) who contributed already to the optimization \emph{PD data set} and known to suffer from disturbed sleep. 
After informed written consent, the patient was recruited and instructed to sleep with the wearable EEG device for two consecutive nights at home. The configuration of the device was switched between the nights to the alternative phase tracking algorithm remotely. The parameters of each algorithm were set to the optimized configuration obtained in the benchmarking experiments. The auditory stimulation was delivered following an ON-OFF stimulation protocol, where stimulations were alternately enabled for a window of 6 s (ON) and immediately thereafter disabled for another 6 s window (OFF) throughout the night. The ON-OFF stimulation protocol enabled for within night analysis of phase targeting performance and SWA enhancement with (ON) and without (OFF) stimulation. For phase targeting analysis, phase at stimulation (or triggers in the OFF cases) were pooled by ON and OFF windows and performance metrics were extracted for each pool. 
The study was conducted according to the Declaration of Helsinki and local legal regulations, and was approved by the institutional research ethics review board (EK ETH 2017-N-67). 

\section{Results}
\label{sec:Results}

\subsection{Benchmarking}

\subsubsection{Phase targeting and stimulation capacity in PD}
\begin{figure}[t]
\centerline{\includegraphics[width=\columnwidth]{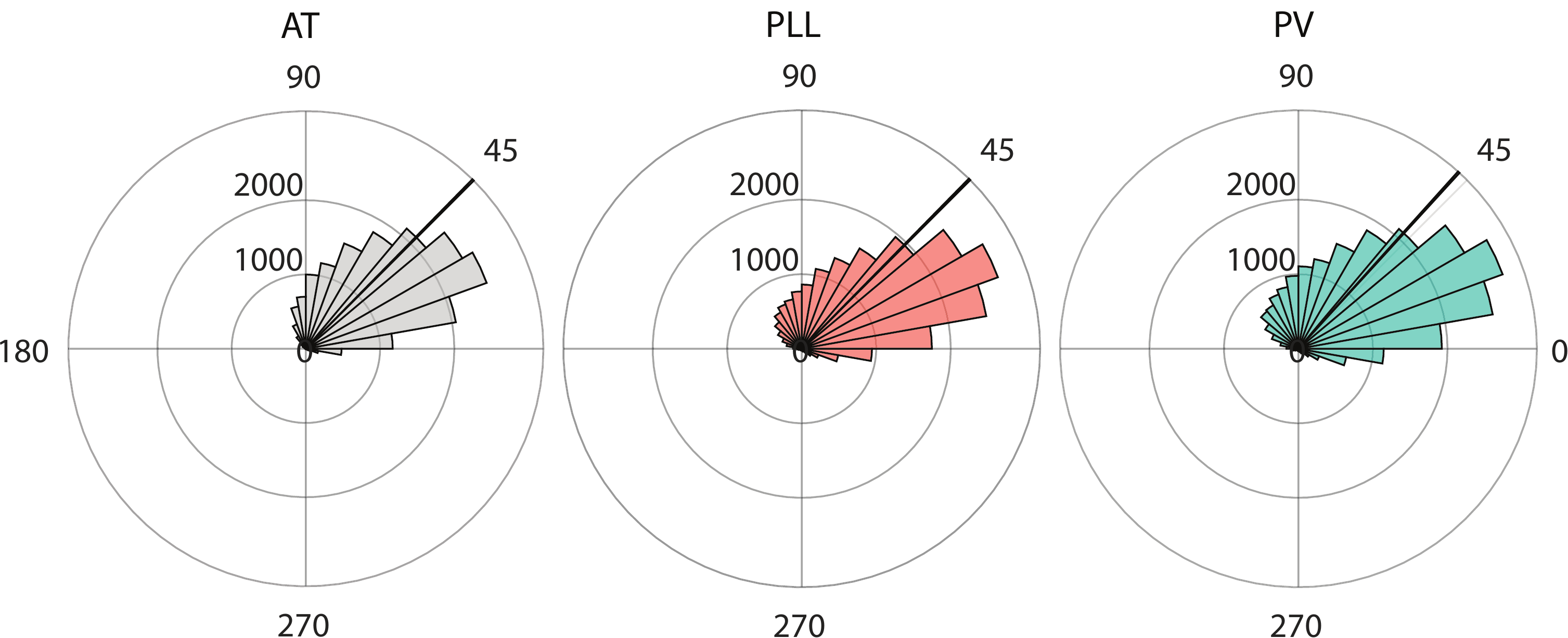}}
\caption{Circular phase distribution at the stimulation trigger event from the {\it PD test set} for the amplitude threshold (AT), phase locked loop (PLL), and phase vocoder (PV). Black solid lines indicate the mean phase vector.}
\label{fig_PDtest_PolarHist}
\end{figure}

All three algorithms showed visually similar phase distributions with the circular mean trending towards the optimization target of 45° (Fig. \ref{fig_PDtest_PolarHist}).  The circular SD  was lower for the AT than for the PLL and the PV (Table \ref{tab:tab_res_all}). The \(\text{PAS}\) and \(\text{PAS}_{\notin\text{UP}}\) were significantly lower than the PLL and the PV. The \(\text{PAS}_{\in\text{UP}}\) and the CMAE$_{\text{45}}$ did not show a significant difference between algorithms (Fig. \ref{fig_PDtest_BoxPlot}(b,d)).

 \subsubsection{Generalization to healthy elderly}
All metrics showed similar results in the {\it elderly test set} when compared to the {\it PD test set}  (Table \ref{tab:tab_res_all}).
 
\subsubsection{Targeting low SW amplitudes}
\begin{figure*} [t]
\centerline{\includegraphics[width=0.85\linewidth]{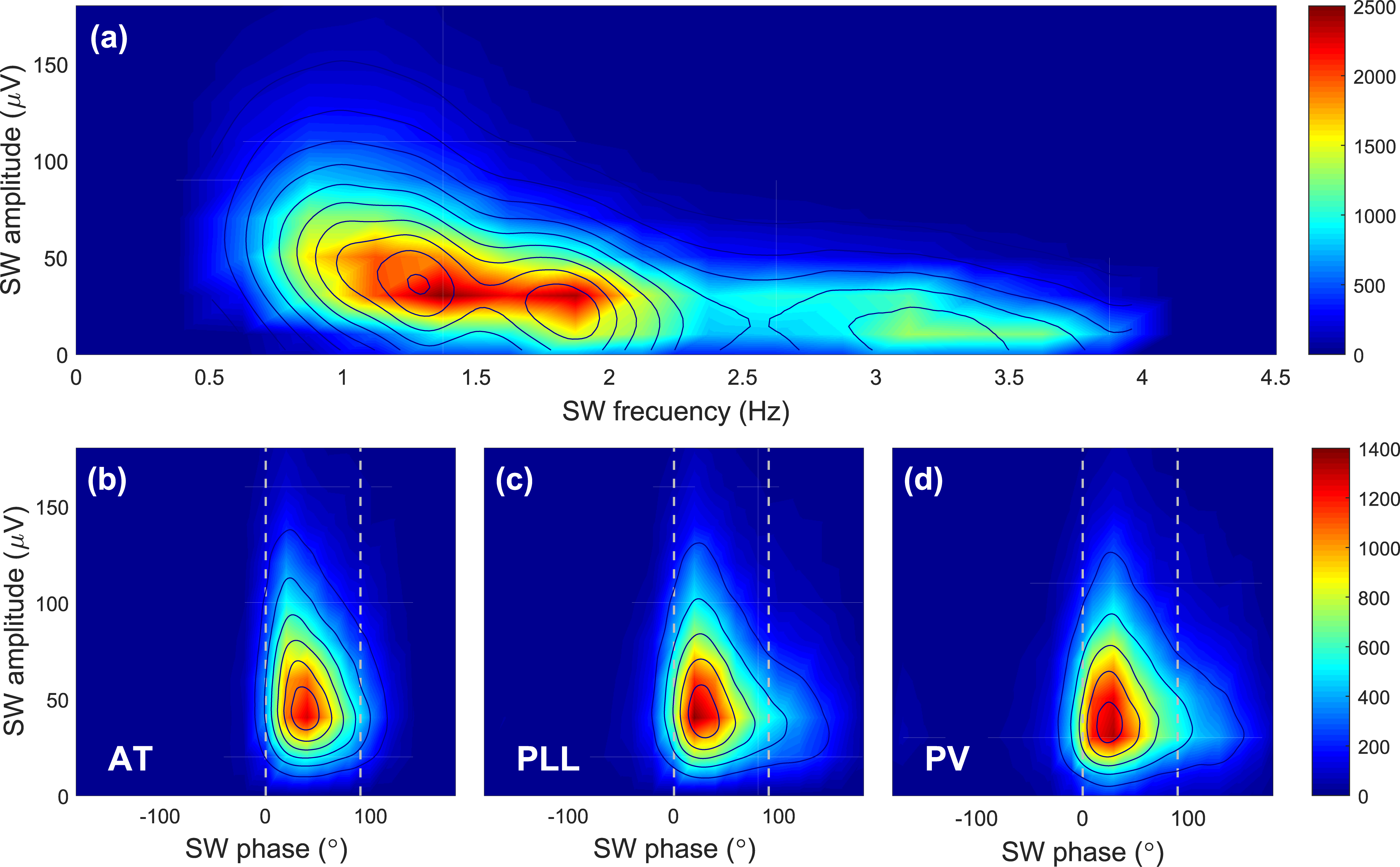}}
\caption{(a) EEG Slow waves (SW) characteristics (frequency and  amplitude) during scored NEM sleep and device NREM sleep and SWA classification of the \emph{PD test set}. SW phase versus SW amplitude at trigger events for the (b) amplitude threshold (AT), (c) phase-locked loop (PLL), and (d) phase vocoder (PV) algorithms.}
\label{fig_PDtest_SWamp_vs_Phase}
\end{figure*}

SW in the {\it PD test set} were characterized by a peak-to-peak amplitude dominating the 20 to 60 \micro V range and a frequency dominating the 1 to 2 Hz range (Fig. \ref{fig_PDtest_SWamp_vs_Phase} (a)). Across algorithms, the triggers were concentrated around  50 \micro V and 45° (Fig. \ref{fig_PDtest_SWamp_vs_Phase} (b-d)). Higher SW amplitudes generated less triggers at SW phases above 45°. 
All algorithms delivered over 70\% of triggers in the desired SW up-phase. The PV showed the highest capacity on targeting low-amplitude SW (74.2\%), whereas the AT only reached 32.3\% (Table \ref{tab:tab_SW}). In comparison to the AT, the phase tracking algorithms targeted more high-amplitude SW.

\subsubsection{Targeting different SW frequencies}
\begin{figure}[t]
\centerline{\includegraphics[width=\columnwidth]{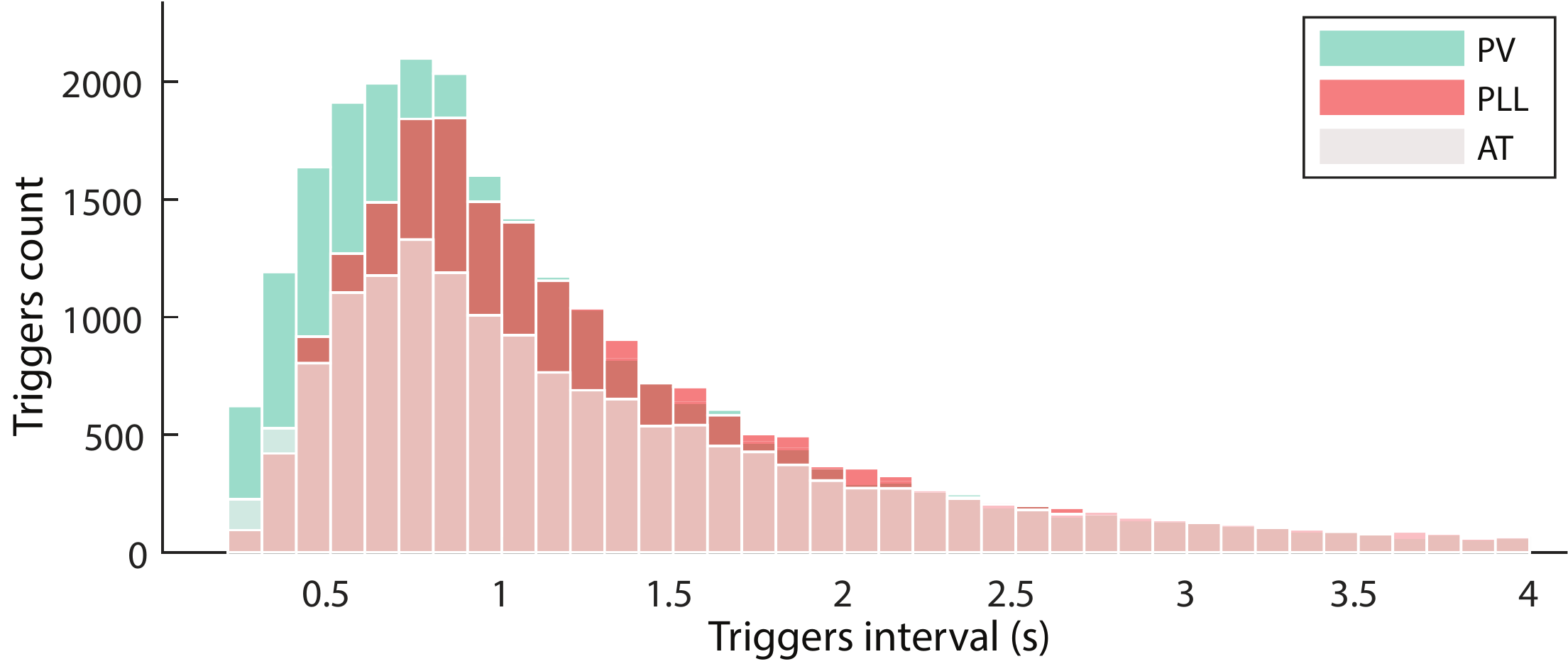}}
\caption{Trigger interval distribution for amplitude threshold (AT), phase-locked loop (PLL), and phase vocoder (PV) algorithms measured from the \emph{PD test set}.}
\label{fig_PDtest_InterToneTime}
\end{figure}

 The PV delivered the highest amount of triggers in the interval range of 0.25 s to 1 s (Fig. \ref{fig_PDtest_InterToneTime}), with the median trigger interval of 0.92 s, slower than the PLL (Table \ref{tab:tab_SW}). The PV targeted more SW at frequencies higher than 1 Hz when compared to the PLL and AT. 

\subsection{Translation to wearable hardware}
The CM and CSD of the phase tracking algorithms in the EEG hardware simulation showed no statistical difference when compared to the {\it PD test set} (Table \ref{tab:tab_res_all}). 
The median number of cycles to complete an algorithm run were 5725 \added{ [Q1 5725, Q3 5761] } for the PLL vs 15476 \added{[Q1 15319, Q3 15544]} for the PV. These resulted in a RCR of  0.02 for the PLL and 0.06 for the PV and an efficiency of 98\% and 94\% respectively.

\subsection{Translation to patients}

Human testing showed that active stimulations have small differences on the CMAE$_{45}$ (Table \ref{tab:tab_on_vs_off}). However, the PLL showed a larger phase shift with respect to the target 45$^{\circ}$ (Table \ref{tab:tab_res_all}). 
When comparing ON vs OFF windows with at least one stimulation,  a significant increase in ON-spectral power during detected NREM sleep was observed in the frequency slow wave range between 0.5 – 4 Hz for the PV and PLL, indicating a response to the auditory stimulation. Within night phase targeting analysis showed comparable performance between ON and OFF stimulation conditions for both phase tracking algorithms (Table \ref{tab:tab_on_vs_off}). 

 \begin{figure}[t]
\centerline{\includegraphics[width=\columnwidth]{Figures/Ferst7.eps}}
\caption{Performance of the amplitude threshold (AT), phase-locked loop (PLL), and phase vocoder (PV) on the {\it PD test set}. (a) Percentage of active stimulation (PAS$_{\text{ALL}}$) of all triggers. (b) PAS in (\(\text{PAS}_{\in\text{UP}}\)) and (c) outside (\(\text{PAS}_{\notin\text{UP}}\)) the up-phase range of slow waves. (d) Circular mean absolute error (CMAE$_{\text{45}}$) of the phase recorded at the stimulation trigger events. The boxes  illustrate the statistical distribution of each performance metric across subjects (outliers not shown). Grey circles represent individual subjects. \(^{*}p <\) 0.05, \(^{**}p <\) 0.001 for paired t-tests.}
\label{fig_PDtest_BoxPlot}
\end{figure}

\begin{table}[t]
  \caption{Results for the benchmarking and the hardware testing across entire data set}
  \centering
  \setlength{\tabcolsep}{3pt}
  \begin{tabular}{l|ccc|ccc|cc|cc}
 \hline 
  
 Performance    & \multicolumn{6}{c|}{Benchmarking (Software)} & \multicolumn{4}{c}{Hardware}  \\ 
     
 Metrics   & \multicolumn{3}{c|}{PD test set} & \multicolumn{3}{c|}{Elderly test set} & \multicolumn{2}{c|}{EEG sim} & \multicolumn{2}{c}{PD subject} \\ 

 & \multicolumn{1}{c}{AT} & \multicolumn{1}{c}{PLL} & \multicolumn{1}{c|}{PV} & \multicolumn{1}{c}{AT} & \multicolumn{1}{c}{PLL} & \multicolumn{1}{c|}{PV} & \multicolumn{1}{c}{PLL} & \multicolumn{1}{c|}{PV} & \multicolumn{1}{c}{PLL} & \multicolumn{1}{c}{PV} \\

   \hline 
   
   &  \multicolumn{3}{c|}{} & \multicolumn{3}{c|}{} & \multicolumn{2}{c|}{} &  \multicolumn{2}{c}{}  \\ 
 
CMAE$_{45}$ ($ ^{\circ}$)  & 0.4  & 0.2  & 2.5    &   1.8  & 2.8  & 2.3       &    7.8   & 3.3    &   14.0  & 1.1 \\
 
 CM  ($ ^{\circ}$) & 45.4 & 45.2 & 47.5 & 43.2 & 42.2 & 47.2 & 37.2 &  48.3 &  31.0 & 43.9\\
 
 CSD ($ ^{\circ}$)   & 32 & 45.9 & 48.6    &   33.8 & 47.6 & 52.7   &    41.4  & 42.6   &    43.4  & 44.4 \\

$\text{PAS}_{\text{ALL}}$ ($\%$)     & 16.7 & 20.4 & 23.3   &   15.3 & 19 & 22.4   &    16.2   & 16.3   &    22.4  & 20.3 \\

$\text{PAS}_{\in\text{UP}}$ ($ \% $) & 14.2  & 14.7 & 16    &   12.7 & 13.2 & 14.3   &    12.8  & 12.6   &    15.9  & 15.1 \\
 
$\text{PAS}_{\notin\text{UP}}$ ($ \% $) & 2.5  & 5.7  & 7.3    &    2.6 & 5.8 & 8.1      &     3.4  & 3.7    &    6.5  & 5.2 \\
   \hline
    \end{tabular}
  \label{tab:tab_res_all}
\end{table}

\begin{table}[t]
  \caption{Algorithms' capacity of trigger different  slow wave characteristics (PD test set)}
  \begin{tabular}{l|ccc}
  \hline
\multicolumn{1}{c}{Metric} &  \multicolumn{1}{c}{AT} & \multicolumn{1}{c}{PLL} & \multicolumn{1}{c}{PV}  \\
\hline
& \multicolumn{3}{c}{} \\ 
Total number of triggers & 18075 & 22442 & 24307 \\
SW up-phase triggers  (\%) & 86.3   &  72.4  & 70.1   \\
Targeted low amp SW    (\%) & 32.3 &  43.5 & 47.2 \\
Targeted high amp SW   (\%) & 74.5 &  81.6 & 81.1 \\
\hline
Median (SD) trigg. interval (s) & 1.07 (0.83)  & 1.06 (0.75)  &  0.92 (0.78)\\
   \hline 
    \end{tabular}
  \label{tab:tab_SW}
\end{table}

\begin{table}[t]
  \caption{Within night results of phase tracking algorithms under stimulation (ON) and non-stimulation (OFF) on human testing.}
  \centering
  \setlength{\tabcolsep}{12pt}
  \begin{tabular}{l|cc|cc}
 \hline
  
 Performance    & \multicolumn{2}{c|}{PLL} & \multicolumn{2}{c}{PV}  \\ 
     
 Metrics        & ON & OFF & ON & OFF \\
 \hline 
   
   &  \multicolumn{2}{c|}{} & \multicolumn{2}{c}{}  \\ 
 
CMAE$_{45}$ ($ ^{\circ}$)  & 14 &   11.7  &   1.1  &  5.3 \\
 
 CM  ($ ^{\circ}$) &  31  &  33.3 &  43.9 &   50.3\\
 
 CSD ($ ^{\circ}$)   & 43.4  &  45.8 &  44.4 & 45.7 \\

$\text{PAS}_{\text{ALL}}$ ($\%$)               & 22.4  & 20.8 & 20.3 &  22.1 \\

$\text{PAS}_{\in\text{UP}}$ ($ \% $)    & 15.9  & 14.6 & 15 & 16.2 \\
 
$\text{PAS}_{\notin\text{UP}}$ ($ \% $) & 6.5   & 6.3  &  5.3 &  5.9 \\
   \hline
    \end{tabular}
  \label{tab:tab_on_vs_off}
\end{table}

\section{Discussion}
\label{sec:Discussion}

We have \added{presented and} benchmarked three algorithms for real-time EEG phase estimation. We highlighted that with optimized algorithm parameters, phase-accurate stimulations can also be delivered in challenging data sets from population known to have reduced SW amplitudes during NREM sleep. \added{We designed the algorithms specifically for wearable hardware  and to achieve better phase tracking performance. We simplified the} PLL approach to be computationally more efficient and reduce the number of parameters to configure. The PV was introduced to estimate the SW phase in real time. This is the first time an algorithm has been designed to track the SW phase across the entire SW frequency band without locking into a fixed center frequency of the EEG. The embedded PV efficiency was 4\% lower than the PLL. We presented for the first time an \added{efficient, phase-accurate,} and fully autonomous phase tracking algorithm embedded in a wearable system for SW  auditory stimulation during sleep.  

SWA is reduced with increasing age and the presence of neurological diseases \cite{Espiritu2008}, making the data from PD participants and older adults investigated in this study a suitable challenge for evaluating real-time SW phase targeting algorithms. To benchmark the algorithms on such data, we performed an optimization on an equal dataset and we introduced novel metrics that normalize the results and focus on the most salient performance information. In the past, studies reported results based on absolute numbers of stimulations  that were dependent on individual sleep characteristics leading to a strong bias towards longer nights. The inclusion of subjects with long continuous NREM sleep stages, made it impossible to compare results across studies or subjects. Our proposed metric based on PAS during NREM sleep overcomes this limitation. PAS was not only used for quantitatively evaluate an algorithm's performance, but also as objective cost function in the optimization process.

With the optimized parameters, the AT and PLL  algorithms showed lower CMAE$_{45}$, CSD and PAS$_{\notin\text{UP}}$ than PV in the {\it PD test set} which was drawn from the same population as the optimization data. However, when tested on the  {\it elderly test set}, which was collected in an independent study, the CMAE$_{\text{45}}$ remained stable with the PV while it increased for the AT and PLL. This suggested that the PV architecture was capable of adapting to individual sleep patterns and characteristics more easily without the need for further parameters' tuning.

Unsurprisingly, the AT algorithm, which did not track phase but only provide an estimate based on amplitudes showed the lowest rate of targeted low amplitude SW and the lowest number of stimuli delivered. In subjects who have limited stimulation opportunities due to reduced sleep and high sleep fragmentation \cite{Lo2014,Petit2004}, such as older adults and PD patients, stimulation effectiveness could be reduced. Phase tracking algorithms had the higher potential to follow SW in real time as they did not rely on a  minimum EEG amplitude condition. The PV excelled in this task and triggered low-amplitude SW in 47.2\% in comparison to PLL (43.5\%) and AT (32.5\%). Together with its capacity to follow the phase at higher frequencies, the PV was the better choice to maximise the delivery of auditory stimuli during the up-phase of SW.

To date, only few studies performed auditory stimulation of SW in populations over 50 years of age. Theses studies observed only small or no SWA enhancement when compared with nights without stimulation \cite{Manuscript2018b,Papalambros2017,Schneider2020SusceptibilityAge}. The circular phase distribution observed in a study carried out in older adults aged 60 to 84 years where a PLL algorithm was used for phase tracking showed a high rate of stimulation outside the desired phase rage (90$\degree$ to 315$\degree$), something that could not be observed when SW were limited to amplitudes larger than 40 \micro V  \cite{Papalambros2017}. In our study, we demonstrated that specifically designed algorithms can avoid stimulations outside the desired range, without compromising performance inside the desired range. As our data sets were limited to PD and healthy older populations, we cannot generalize these findings to healthy young and other populations. Thus, to establish an estimation of the robustness of the PV and simplified PLL, further studies on extended populations remains to be conducted.

When implemented on a wearable platform, the computational complexity of the phase tracking algorithms had only marginal effect on autonomy, which was above 20 hours. This demonstrated the suitability of the proposed algorithms for phase-accurate auditory stimulation in wearable and low-power devices.   

Our analysis on prospectively collected stimulated sleep EEG on a PD patient who showed within night SWA enhancement in an ON-OFF stimulation approach, revealed that the phase tracking is not affected by the auditory stimulation that might cause phase shifts in the EEG. This opens the way for deployment of the algorithms in larger studies to \added{further validate the presented algorithms and investigate} the effect of the auditory stimulation on sleep and human health. 

\section{Conclusion}
\label{sec:Conclusion}
This work introduced for the first time a PV based algorithm to autonomously track the EEG phase  over a broad frequency range in real time. We demonstrated that the PV provides a higher capacity for targeting low SW amplitudes and frequencies above 1 Hz when compared to the PLL and AT detection algorithms. Due to the relevance of auditory stimulation for populations whose sleep is hallmarked by reduced SW amplitudes, this new technique has the potential to accurately target more SW and consequently increase the desired SWA enhancement effect.

\section*{Acknowledgements}
\label{sec:Ack}
We are grateful to  all participants who contributed their data. We thank R. Büchi, N. Demarmels, S. Huwiler, G. Hoppeler, J. Kurz, E. Silberschmidt, and L. Kämpf who performed the recruitment and data collection; M. Furrer and S. Cortesi who assisted with the data collection  for the hardware simulation experiments; S. Fattinger, E. Werth, and R. Poryazova who assisted with sleep scoring; and A. Gómez and P. Di Giulio who provided valuable feedback on this manuscript. We thank all SleepLoop consortium members who contributed to uncountable discussions and feedback on this work. 

\textit{Conflict of interest:}  R. Huber, C. Baumann, and W. Karlen are founders and shareholders of Tosoo AG.

\printbibliography

\end{document}